\documentclass[aps,pre,preprint,superscriptaddress]{revtex4-1}

\begin{document}


\title{Effects of Counterions on Nano-confined DNA Thin Films}


\author{Nupur Biswas}\affiliation{Applied Material Science Division, Saha Institute of Nuclear Physics, 1/AF Bidhannagar, Kolkata 700064, India}\altaffiliation{Current address: Department of Physics, Indian Institute of Science, Bangalore 560012, India}
\author{Sreeja Chakraborty}\affiliation{Chemical Science Division, Saha Institute of Nuclear Physics, 1/AF Bidhannagar, Kolkata 700064, India}
\author{Alokmay Datta}\email{alokmay.datta@saha.ac.in}\affiliation{Surface Physics and Material Science Division, Saha Institute of Nuclear Physics, 1/AF Bidhannagar, Kolkata 700064, India}
\author{Munna Sarkar}\affiliation{Chemical Science Division, Saha Institute of Nuclear Physics, 1/AF Bidhannagar, Kolkata 700064, India}
\author{Mrinmay K. Mukhopadhyay}\affiliation{Surface Physics and Material Science Division, Saha Institute of Nuclear Physics, 1/AF Bidhannagar, Kolkata 700064, India}
\author{Mrinal K. Bera}\affiliation{Center for Advanced Radiation Sources, University of Chicago, Chicago, Illinois 60637, USA}
\author{Hideki Seto}\affiliation{KENS $\&$ CMRC, Institute of Materials Structure Science, High Energy Accelerator Research Organization, Tsukuba 305-0801, Japan}



\date{\today}

\begin{abstract}
DNA molecules spin-coated on amorphous quartz substrates are shown to form stable films. Electron density profiles (EDPs) along the film depth show that film prepared from aqueous solution of DNA exhibits layering of the molecules in three stacks parallel to the substrate whereas film prepared from counterion added solution does not have layering but have smaller thickness and enhanced surface roughness, although both films have `liquid-like' height-height correlations. We explain these results by a model of film of a `liquid' comprising of rod-like molecules where the counterion concentration in the liquid determines the rod length.

\end{abstract}

\pacs{}

\maketitle

\section{Introduction}
DNA or deoxyribonucleic acid is a macromolecule having a diameter of about 22-26 {\AA} \cite{JMB81}. In aqueous medium, specially in presence of cations the acid group dissociates to generate a macromolecular polyanion with a persistence length of at least $\sim$ 500 {\AA} \cite{PNAS-94-6185-1997}. The counterions, provided by dissolved inorganic or organic salts (buffers), screen the negative charge and stabilize the DNA as dispersion in aqueous medium. As expected from dilute solutions and also as confirmed from simulation studies, the DNA molecules are dispersed as gyration spheroids with a mean radius increasing with decreasing counterion concentration \cite{JPCB-109-10458-2005,JMB-267-299-1997,Biopolymers-31-1471-1991}.

However, it is well known that (a) in biologically meaningful conditions, such as in the cellular environment, DNA molecules exist in a confined space of micrometer diameter, crowded with counterions and smaller molecules \cite{Biochimie-90-1040-2008}, and (b) in response to screened Coulomb forces and/or short-range forces (the most well-known being the `depletion force' \cite{JPolySc-33-183-1958}) DNA molecules may organize into liquid crystalline or crystalline phases in extra-cellular bulk \cite{Livolant}. Response of DNA towards this crowding is also influenced by the confining geometry \cite{NanoLett-11-5047-2011,JCP-128-225109-2008}.

In a previous work we have shown that a DNA-PEG-NaCl system that forms a monophase aqueous system in bulk phase segregates spontaneously when confined within a droplet of micrometer length scale with the DNA molecules forming a layer at the boundary, leaving the interior of the droplet severely depleted. It was also found that this outer layer of DNA molecules is birefringent showing a molecular self-organization to an ordered structure. Most importantly, this segregation and ordering was clearly correlated with confinement, becoming slower with increase in the droplet diameter indicating a critical size above which the process will be essentially non-existent \cite{CPL-539-157-2012}.

However, the evolution of the ordering of DNA molecules in the segregated layer is essentially a nanometer-scale phenomenon \cite{JCP-128-225109-2008} and the above study, which concentrated on the droplet as a whole and thus was at micrometer length scales, was unable to probe this level of self-organization. This is important for several reasons. First, both simple and complex liquids, when confined to nanometer length scales, especially at an interface, can spontaneously form one-dimensionally and two-dimensionally \cite{PRL-82-2326-1999,PRE-63-021205-2001,PRL-96-096107-2006,Macro-33-3478-2000} ordered structures and similar structures in DNA under confinement may provide spatially organized sites for molecular and supra-molecular bonds. Secondly, the height-height correlation at the surface or interface of such structures maybe either `solid-like' or `liquid-like', and this has a bearing on their mechanical properties. Also, one dimensionally nano-confined films act as mimics of biologically relevant immobile structures that provide steric constraints as well as confined environment. Unfortunately, the nanoconfined liquids studied so far, are mostly neutral \cite{PRL-83-564-1999}. In particular, to our knowledge, no such study has been carried out on polyelectrolytes or charged macromolecules, although such studies are specifically important for understanding biologically interesting layered structures.

In the present study, we have tried to address the above issues by studying the in-plane and out-of-plane structure and morphology of nano-confined films of DNA in the pristine state and in presence of counterions provided by buffer molecules. We have used Atomic Force Microscopy for studying the surface morphology, X-ray reflectivity to extract the electron density profile (EDP) along the film depth and X-ray diffuse scattering to find the height-height correlation on the film surface, which decides whether the films are `solid-like' or `liquid-like'. The spin-coated films in our studies may serve as planar models to understand the nanoscale self-organization in the cellular space. The questions we are trying to answer are (a) whether nano-confined DNA molecules form layers spontaneously ? (b) What role do the counterions play in this layering ? and (c) whether the films are `solid-like' or `liquid-like' in their height-height correlations ?

\section{Experimental Details}
Polymerized calf thymus DNA (Sisco Research Laboratory, India) in triple distilled water formed the pristine stock solution. The absorbance ratio A$_{260}$/A$_{280}$  of the solution at 260 nm and 280 nm being in the range 1.8 $<$ A$_{260}$/A$_{280}$ $<$ 1.9, indicated that no further deproteinization of the solution was necessary. Concentration of the stock solution in terms of nucleotide, assuming $\varepsilon_{260}$ = 6600 M$^{-1}$cm$^{-1}$, was found to be 1.8mM. The stock solution was diluted to the desired concentration of 800 $\mu$M in triple distilled water and was used to prepare `pristine' films. 10mM of sodium cacodylate (Merck, Germany) solution in triple distilled water was adjusted to the desired pH of 6.7 with hydrochloric acid and was used as the buffer. This solution was used to prepare the `buffered' films with the 800$\mu$M solution of DNA. The buffer concentration was maintained well below the critical monovalent counterion concentration of $\sim$ 500mM, a condition which is required for complete neutralization of DNAs \cite{PNAS-94-6185-1997}. Also use of high salt concentration (500mM) leaves excess salt crystals over the film surface and consequently prevents us from probing the layered structures \cite{AIP-1447-189-2012}.

Films were prepared by spin-coating the solution at 4000 rpm on amorphous fused quartz substrates at ambient condition using a spin-coater (Headway Research Inc., USA). Before spin-coating the fused quartz (Alfa Aesar, USA) substrates were cleaned and hydrophilized by boiling in 5:1:1 H$_2$O:H$_2$O$_2$:NH$_4$OH solution for 10 minutes, followed by sonication in acetone and ethanol respectively. The substrates were then rinsed by Millipore water (resistivity $\sim$18.2 M$\Omega$ cm) and subsequently water was removed by spinning the substrate at high speed (4000 rpm). Henceforth, for the sake of convenience, films obtained from pristine and buffer-added solutions would be called `pristine' and `buffered' respectively.

For extracting out-of-plane information we have recorded specular X-ray reflectivity (XRR) profiles of these thin films. This is a well established technique for investigating layered systems. In XRR we measure $\vec{q}$ which is the difference in momentum between the incident and scattered beam. For specular condition momentum transfer occurs only in out-of-plane (Z) direction ($\vec{q} \equiv (q_x=0, q_y=0, q_z=(4\pi/\lambda)sin\theta_i\neq 0$), $\theta_i$ = angle of incidence) i.e. perpendicular to the film plane which is also the confining direction. Since $\vec{q}$ has dependency both on incidence angle and wavelength of the X-ray beam, XRR data can be recorded in angle dispersive and as well as in energy dispersive mode, respectively \cite{JPhysD-39-R461-2006}. We have followed angle dispersive mode and recorded the intensity of reflected X-ray beams varying the angle of incidence with an angular step size of 5 millidegree at the Indian Beamline (BL-18B) at Photon Factory, High Energy Accelerator Research Organization (KEK), Japan using X-ray of wavelength ($\lambda$) 1.08421{\AA}. This reflected intensity is the resultant of interference of X-rays reflected from different interfaces of the film and thus contains information of those interfaces.

On the other hand, in transverse diffuse scattering, we measure off-specular intensity of the scattered beam for in-plane information \cite{RepProgPhys-63-1725-2000}. Here also we measure intensity by varying angle of incidence i.e. by rocking the sample but keeping the detector fixed. Here we maintain $q_x\neq q_y\neq q_z\neq0$. Since X-ray beam was incident along Y direction and source slit dimension was 0.1mm and 2mm in vertical and horizontal direction, respectively, i.e. quite large in out-of-the scattering plane or X direction, we can easily assume this scattering geometry effectively integrates out the intensity along $q_x$ direction leaving intensity as a function of $q_y$ only. Since off-specular scattering is dominated by in-plane scattering from the sample surface, this beam carries information of surface height distribution. We recorded transverse diffuse scattering data with step size of 2 millidegree for three different angular positions of detector and thus in-plane information at three different depths of the sample was collected. The sample was kept in nitrogen atmosphere to avoid radiation damage.

Atomic Force Microscope (AFM) images were recorded at tapping mode using Nanonics MultiView1000 with Au coated cantilevered AFM probes of glass of tip diameter $\sim$ 20nm. The scan size was chosen 5$\mu$m $\times$ 5$\mu$m to probe the long range characteristics. The images were analyzed using WSxM software \cite{WSxM}.

\section{Results and Discussions}
\subsection{Layering and non-layering}
X-ray reflectivity profiles are outcome of scattering of an electromagnetic wave by matter in specular direction. It is analyzed by different formalism varying the extent of approximations in perturbation of electromagnetic wave by matter \cite{Daillant,PhysRep-257-223-1995}. We have used the Distorted Wave Born Approximation (DWBA) \cite{jkbasu}, which only requires an ansatz of the average electron density of the film and uses the exact (Maxwell's) wavefunctions of the beam scattered from the film to find out the electron densities of different `layers' of the film. The thickness of the layers are decided by the spatial resolution along Z, which in turn is given by the maximum value of the momentum transfer $q_z$ up to which Kiessig (interference) fringes are observed. This technique is efficient for smaller variations and relatively unknown compositions, and has been used extensively for detecting layering in nano-confined liquids \cite{EPL96,PRB,Macro2}. The EDP shown in Figure 1b was obtained by further convoluting it with air-film and film-substrate interfacial widths within the resolution limits, where these widths were also obtained from DWBA.

Let us first discuss the pristine film. DWBA fit (Figure 1a) gives the thickness of this film as $\sim$ 78{\AA}, corresponding to three layers of DNA lying on their sides parallel to the substrate surface, on top of each other. The widths of the air-film and film-substrate interfaces, $\sigma_{af}$ and $\sigma_{fs}$, are 5.0{\AA} and 6.9{\AA}, respectively (Table 1). Since the $\sigma_{af}$ of bare quartz is $\sim$ 7{\AA}, this indicates the excellent flatness and smoothness of the film, underscoring its stability. The value of $\sigma_{af}$ is consistent with the value of r.m.s. roughness ($\sigma_{rms}$) of the film surface, as obtained from AFM (Figure 1c and Table 1).

The electron density profile (EDP) extracted from this fit (Figure 1b) shows formation of three distinct density oscillations characteristic of layering with a periodicity of $\sim$ 26{\AA}, nearly the diameter of the DNA molecule, rather than the polymer radius of gyration \cite{PRB}. The order parameter for layering, $\delta = \rho_{max} - \rho_{min}$, where $\rho_{max}(\rho_{min})$ is the average maximum(minimum) electron density in the film EDP comes out to be 0.032 e{\AA}$^{-3}$, and the $\sigma_{int}$ i.e. the interfacial width between the layers is negligible. This confirms the model of this film being that of a stack of three layers of pristine DNA molecules lying on their sides and aligning themselves parallel to the hydrophilic substrate, similar to short-chain polymers and simple liquids \cite{Langmuir-17-4021-2001}. Such confinement induced layering of DNA molecules is also in accordance with previous reports of enhancement of asymmetric shape of DNA molecules confined in nanoslits \cite{Macro-45-2920-2012}.

Let us now look at the buffered film. The DWBA fit in Figure 1a yields the EDP in Figure 1b. The values of $\sigma_{af}$ and $\sigma_{fs}$ are 9.0{\AA} and 8{\AA}, respectively. Though the film quality is still very good but this increase in interfacial widths, especially $\sigma_{af}$, is partially explained from the AFM topographical image in Figure 1d. This shows formation of nanometer-sized islands that takes the r.m.s. surface roughness to this higher value. The probable origin of these islands will be discussed in next section. Nevertheless, the overall r.m.s. roughness of the film surface is $\sim$ 19 {\AA}, while the roughness between the islands is $\sim$ 9 {\AA} which is shown in Table 1.

The most interesting aspects of the EDP of the buffered film are its thickness $\sim$ 52{\AA}, which corresponds to the thickness of a stack of two DNA molecules aligned parallel to the substrate, lying on their sides, and the absence of any density oscillation due to layering. This suggests (a) reduction of adhesion and/or cohesion of DNA, and (b) some form of intermolecular `diffusion' or `overlap' that reduces the order parameter $\delta$ to near-zero value.

\subsection{`Liquids' of rods with adjustable lengths}
We have used surface correlation function to characterize the in-plane morphology of the films. The correlation between the heights of two positions i.e. between the positions of two scattering centers determines the extent of interference of the beam undergoing in-plane scattering \cite{Tolan}. Following this principle we have extracted height-difference correlation from transverse diffuse scattering data. Under DWBA the diffuse scattering cross section is given by \cite{PRB-38-2297-1988},
\begin{equation}\label{Eqn:DSC}
    \frac{d\sigma}{d\Omega} \sim \mid T(\vec{k_1})\mid^2 \mid T(\vec{k_2})\mid^2 S(\vec{q_t})
\end{equation} where $T(\vec{k_1})$ and $T(\vec{k_2})$ denote transmission coefficient for incident and outgoing wavevectors $\vec{k_1}$ and $\vec{k_2}$, respectively and structure factor $S(\vec{q_t})$ for transmitted wavevector $\vec{q_t}$ is given by,
\begin{equation}\label{Eqn:S}
    S(\vec{q_t})= \frac{exp\{-[(q_z^t)^2+(q_z^{t*})^2]\sigma^2/2\}}{\mid q_z^t \mid^2}\int\int_{S_0}dX dY (e^{\mid q_z^t\mid^2(\sigma^2-0.5g(X,Y))}-1)e^{-i(q_xX+q_yY)}
\end{equation}
Eq. \ref{Eqn:S} contains the height-difference correlation function $g(X,Y)$ which in polar coordinate for an isotropic Gaussian rough surface is given by,
\begin{equation}\label{Eqn:Corr}
    g(r) = <[h(r_0+r)-h(r_0)]^2>
\end{equation}
where $h(r)$ denotes the height at any point $r$ and $<>$ denotes ensemble average over all possible surface configurations. Our transverse diffuse scattering data of the films taken at different detector angles, i.e. probing at different depth of film, fits well with the `self-affine liquid' (SALiq) correlation function \cite{PRL-82-4675-1999},
\begin{equation}\label{Eqn:SALiq}
    g(r)=\{2\sigma^2+B[\gamma_E + ln(\frac{\kappa r}{2})]\}\{1-exp[-(r/\xi)^{2\alpha}]\}
\end{equation}
where $\sigma$= rms roughness, $B=k_BT/\pi\gamma$, ($\gamma$=interfacial tension, $T$=absolute temperature) $\gamma_E$=Euler constant, $\kappa$=lower cut-off wave-vector corresponding to mass/density fluctuation ($\kappa^2=g\triangle\rho/\gamma$, $g$=acceleration due to gravity, $\triangle\rho$=density fluctuation), $\xi$=correlation length, $\alpha$=Hurst exponent. These fits (Figure 2) show that both of these films have `liquid-like' correlation.
Table 2 shows that the roughness $\sigma$ obtained from diffuse scattering at three incident angles match with the $\sigma_{af}$, obtained from X-ray reflectivity and $\sigma_{rms}$, obtained from AFM (Table 1) respectively.

With increasing angle (2$\theta$) of the detector position, the correlation length $\xi$ gradually decreases, consistent with the fact that at higher angle the illuminated area decreases.

Decrease in the value of Hurst exponent ($\alpha$), indicating a more zig-zag surface, is consistent with an enhancement of roughness with buffering. Enhanced value of surface tension ($\gamma$) with increase of counterion concentration too, can be attributed to the enhanced roughness of buffered film \cite{ARMR-38-71-2008}.

At the same time we observe that lower cut-off wavevector ($\kappa$) has decreased significantly from pristine to buffer film, associated with decrease of $\triangle\rho$. Decrease in the value of $\triangle\rho$ in buffer added film and its constant value with various detector angle, i.e. at different depth of film implies uniform distribution of molecules along the film depth, consistent with the absence of layered structure as obtained from XRR data.

Thus we observe (a) pristine and buffered DNA can be spin-coated to form stable films that show `liquid-like' height-height correlations at larger length scales, (b) pristine film consists of layers of DNA with the long-axes of the molecules lying parallel to the substrate surface while the surface roughness is very small and (c) a small buffer (counterion) concentration both destroys the layering and enhances the roughness.
Based on this information we visualize this system as a `frozen' state of vibrating rigid rods.

The question regarding the adhesion of the pristine film to the hydrophilic substrate has been answered by recent simulation results \cite{Macro-44-1707-2011} on polyelectrolyte film attachment to such substrates. With the same view, we propose that the hydronium (H$_3$O$^+$) counterions to the phosphatic negative charges located on the DNA molecules provide attachment to the hydroxyl-terminated hydrophilic substrate through short-range interactions such as hydrogen bonds, over and above the long-range but weak, screened Coulomb attraction, and also that these forces align the DNA persistent length `rods' parallel to the substrate surface. Within the film this lateral alignment is more favourable as besides H-bonding, the self-avoidance is also enhanced by reduced screening of Coulomb interaction \cite{PRL-99-058302-2007}. Figure 3a and 3b shows the schematic of the film structure.

\section{Conclusions}
We have observed within a nano-confined thin film, in a quasi-equilibrium condition, by rearranging themselves DNA molecules form a confined charged liquid. In pristine film they behave like long semi-flexible rods and form layers quite like to simple liquid molecules. Addition of counterions neutralize their charge and make the rods shorter and hence increases the orientational entropy that destroys the layered structure. Addition of counterions thus effectively reduce the `rod' length and make them `soft' on a larger length scale.

Our results show the importance of confinement towards the molecular arrangement of DNAs. Hence role of confining space demands attention in other cell-mimicking systems. It also shows a minute tuning of counterion concentrations can change the arrangement of DNAs and creates an opportunity of controlling the performance of nanofluidic devices \cite {AnalChem-80-2326-2008,PRL-94-196101-2005} where DNA molecules are made confined within nanoslits.

\begin{acknowledgments}
We would like to acknowledge Heiwa-Nakajima Foundation, Japan for providing financial support and Department of Science and Technology, Government of India for sponsoring Indian beamline project at Photon Factory, KEK, Japan. Authors N. B. and S. C. thank Council of Scientific and Industrial Research (CSIR), Government of India and Director, SINP for their research fellowships.
\end{acknowledgments}

{}

\newpage
\underline{\bf TABLE 1}: Parameters obtained from X-Ray Reflectivity and AFM\\

\begin{table}[!h]
\begin{center}
 \caption{Parameters obtained from X-Ray Reflectivity and AFM}\label{Tbl:XRR-Para}
 \begin{tabular}{ccccccc}
 Film & $\sigma_{af}$ ({\AA}) & $\sigma_{fs}$ ({\AA}) & $d$ ({\AA}) & $\sigma_{rms}$ ({\AA})\\
 \hline
 Pristine & 5.0 & 6.9 & 78 & 6.6\\
 Buffered & 9.0 & 8.0 & 52 & 9.0\\
 \end{tabular}
 \end{center}
 \end{table}

\newpage
\underline{\bf TABLE 2}: Parameters obtained from diffuse scattering fit\\

\begin{table*}[!h]
\begin{center}
 \caption{Parameters obtained from diffuse scattering fit}\label{Tbl:DiffusePara}
 \begin{tabular}{cccccccc}
  \hline
 Film & 2$\theta$ & $\sigma$ & $\xi$ & $\alpha$ & $\gamma(10^{-12})$ & $\kappa(10^{-7})$ & $\triangle\rho(10^{-26})^*$\\
 & & ({\AA})& ({\AA}) &  & (N/{\AA}) & ({\AA}$^{-1}$) & (kg/{\AA}$^3$)\\
 \hline
 Pristine & 1.2$^\circ$ & 4.85 & 4000 & 0.13 & 2.03 & 5.01 & 5.2\\
 & 1.5$^\circ$ & 4.5 & 3500 & 0.189 & 3.091 & 6.0 & 11.35\\
 & 1.8$^\circ$ & 5.71 & 3068 & 0.193 & 2.792 & 6.53 & 12.14\\
 \hline
 Buffered & 1.2$^\circ$ & 8.5 & 3100 & 0.072 & 5.019 & 2.0 & 2.05\\
 & 1.5$^\circ$ & 7.9 & 3000	& 0.118	& 5.14	& 2.10 & 2.31\\
 & 1.8$^\circ$ & 8.15 & 2900 & 0.120 & 5.036 & 2.0 & 2.06\\
 \end{tabular}
 \end{center}
 $^*$\emph{Calculated from $\triangle\rho=\kappa^2\gamma/g$ using fitted values of $\kappa$ and $\gamma$}
 \end{table*}

\newpage
\underline{\bf FIGURE CAPTIONS}\\

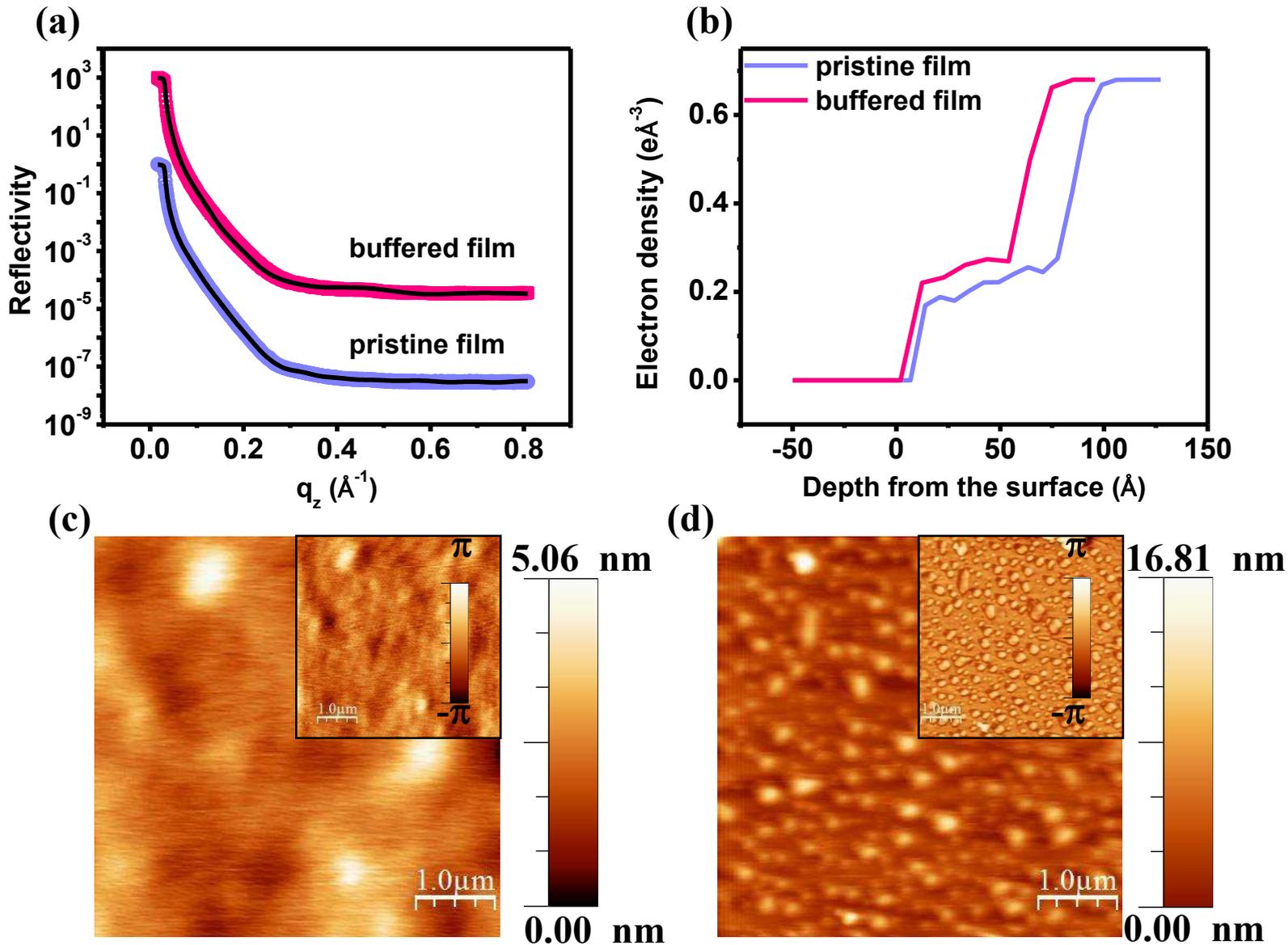
Figure 1: Pristine and buffered DNA films: (a) X-ray reflectivity profiles of pristine and buffered DNA films, fit with DWBA formalism (up shifted for clarity). Symbols: experimental data points; line: fit, (b) Corresponding EDPs, (c) AFM image of pristine film (scan size 5$\mu$m $\times$ 5$\mu$m), (d) AFM image of buffered film (scan size 5$\mu$m $\times$ 5$\mu$m). Insets of (c) and (d) show corresponding phase images.\\

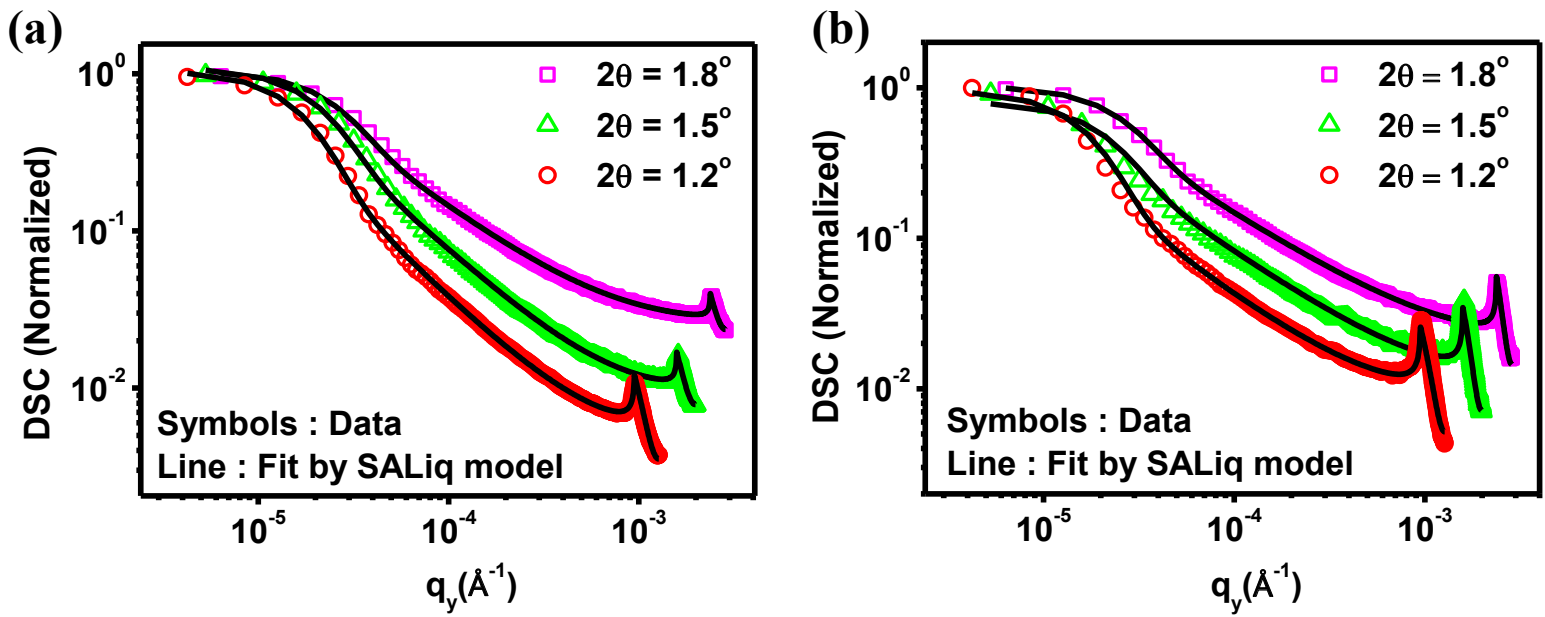
Figure 2: Transverse diffuse scattering data taken at different detector angles (2$\theta$): (a) pristine film and (b) buffered DNA film. DSC stands for Differential Scattering Cross-section. The symbols denote data points whereas lines denote fit by self-affine liquid model under DWBA scheme.\\

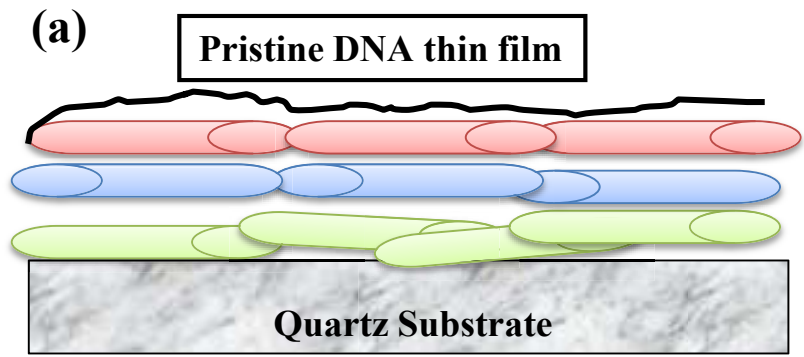
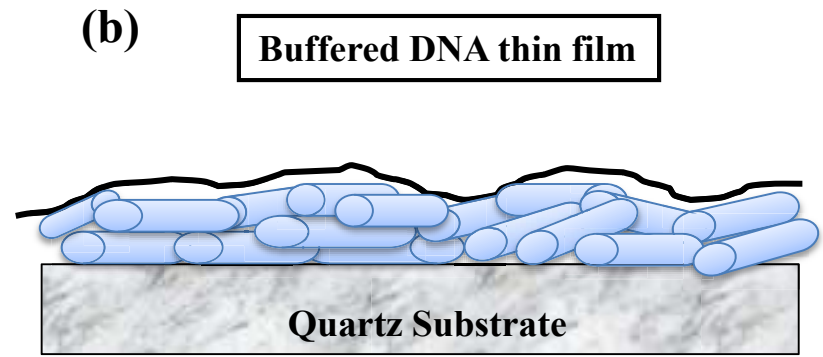
Figure 3: Schematic structure of pristine (a) and buffered (b) film based on our model.\\

\end{document}